\documentclass[a4paper]{article}

\usepackage{INTERSPEECH2021}
\usepackage{amsmath,graphicx}
\usepackage{cite}
\graphicspath{{Figs/}}
\usepackage{float}
\usepackage{graphicx}
\usepackage{epsf}
\usepackage{array}
\usepackage{makecell}
\usepackage{amssymb}
\usepackage{lineno}
\usepackage{fixmath}
\usepackage{hyperref}
\usepackage{float}
\usepackage{caption}
\usepackage{subcaption}
\usepackage{tikz}

\newcolumntype{P}[1]{>{\centering\arraybackslash}p{#1}}
\newsavebox{\foobox}
\newcommand{\slantbox}[2][0]{\mbox{%
		\sbox{\foobox}{#2}%
		\hskip\wd\foobox
		\pdfsave
		\pdfsetmatrix{1 0 #1 1}%
		\llap{\usebox{\foobox}}%
		\pdfrestore
}}
\newcommand\unslant[2][-.25]{\slantbox[#1]{$#2$}}

\title{Nonlinear Acoustic Echo Cancellation with Deep Learning}
\name{Amir Ivry \qquad Israel Cohen \qquad Baruch Berdugo \thanks{This research was supported by the Pazy Research Foundation and the ISF-NSFC joint research program (grant No. 2514/17). The authors thank Stem Audio for providing equipment and technical guidance.}}
\address{Andrew and Erna Viterbi Faculty of Electrical and Computer Engineering \\
	Technion -- Israel Institute of Technology, Technion City, Haifa 3200003, Israel}
\email{sivry@campus.technion.ac.il, icohen@ee.technion.ac.il, bbaruch@technion.ac.il}

\begin{document}
 \sloppy         

\maketitle
\begin{abstract}
	We propose a nonlinear acoustic echo cancellation system, which aims to model the echo path from the far-end signal to the near-end microphone in two parts. Inspired by the physical behavior of modern hands-free devices, we first introduce a novel neural network architecture that is specifically designed to model the nonlinear distortions these devices induce between receiving and playing the far-end signal. To account for variations between devices, we construct this network with trainable memory length and nonlinear activation functions that are not parameterized in advance, but are rather optimized during the training stage using the training data. Second, the network is succeeded by a standard adaptive linear filter that constantly tracks the echo path between the loudspeaker output and the microphone. During training, the network and filter are jointly optimized to learn the network parameters. This system requires 17 thousand parameters that consume 500 Million floating-point operations per second and 40 Kilo-bytes of memory. It also satisfies hands-free communication timing requirements on a standard neural processor, which renders it adequate for embedding on hands-free communication devices. Using 280 hours of real and synthetic data, experiments show advantageous performance compared to competing methods.
\end{abstract}
\noindent\textbf{Index Terms}: Nonlinear acoustic echo cancellation, deep learning, hands-free communication, on-device implementation

\section{Introduction}\label{intro}
Hands-free communication often involves a conversation between two speakers located at near-end and far-end points. The near-end microphone captures the desired-speech signal and two interfering signals: echo produced by a loudspeaker playing the far-end signal, and background noises. The acoustic coupling between the loudspeaker output and the microphone may lead to degraded speech intelligibility in the far-end due to echo presence \cite{sondhi1995stereophonic}. This problem prompted numerous studies regarding acoustic echo cancellation (AEC) systems that aim to remove echo and preserve the near-end speech \cite{benesty2001advances}. In recent years, however, miniaturization of electronic components in hands-free devices, e.g., smart phones, smart speakers, and wearable devices, caused non-negligible nonlinear (NL) distortions in the echo path between the far-end signal and the loudspeaker output \cite{birkett1995limitations}.Consequently, AEC systems that assume an echo path that is linear often fail in practice \cite{mossi2010assessment}. 

To mitigate this mismatch, various nonlinear acoustic echo cancellation (NLAEC) approaches were proposed to identify the NL echo path. The Volterra series showed success in modeling systems with weak nonlinearities and memory using NL basis functions, while often requiring high computational complexity \cite{guerin2003nonlinear}. A simplified version is given by the block-oriented Hammerstein and Wiener models, which describe NL systems without memory and linear systems with memory \cite{scarpiniti2011comparison}. Also, adaptive functional link filters \cite{comminiello2013functional}, Bayesian state-space modeling \cite{malik2012state}, and kernel-based methods \cite{van2016split} are commonly used for NLAEC. Avargel and Cohen considered this problem from a time-frequency point-of-view and applied multiplicative function approximation \cite{avargel2008nonlinear}, sub-band adaptive filtering \cite{avargel2009adaptive}, and an efficient Volttera series modeling using cross-band terms \cite{avargel2009modeling}, \cite{avargelBookChapter}.
Neural networks (NNs) provide an alternative framework for a more accurate NL modeling compared to classic approaches \cite{birkett1995acoustic}, \cite{rabaa1998acoustic}, \cite{janczak2004identification}, \cite{zhang2017recursive}. For instance, Malek and Koldovsky \cite{malek2016hammerstein} estimated the NL echo path with a fully-connected NN (FCNN) that assumes the Hammerstein model, followed by an adaptive linear filter to track the acoustic path. Recently, Halimeh et al. \cite{halimeh2019neural} constructed an FCNN that assumes the Wiener-Hammerstein model and captures both the NL and linear echo paths.

Despite showing promising results, the performance of these methods is still challenging in real-life scenarios, which may be associated with two of their attributes. First, these models are not accurately designed according to the physical behavior of distortions that modern hands-free devices apply to the far-end signal. Second, they are mostly parametric, i.e., they require that memory lengths and NL basis functions are predetermined. E.g., in \cite{guerin2003nonlinear}, \cite{scarpiniti2011comparison}, the presented models assume a given number of memory taps, and in \cite{malek2016hammerstein}, \cite{halimeh2019neural}, fixed NL activation functions are employed inside the NN. These drawbacks may produce sub-optimal solutions in real setups.

To address these two gaps, we make two contributions that are inspired by the physical behavior of modern hands-free devices. We first introduce a novel NN architecture that is specifically designed to model the distortions these devices induce between receiving and playing the far-end signal. Second, we construct this NN with trainable memory length and NL activation functions that are not parameterized in advance, but are rather optimized during the training stage based on the training data. The NN output is inserted into a standard adaptive linear filter that constantly tracks the acoustic path from the loudspeaker output to the microphone. The end-to-end system, from the input of the NN to the output of the linear filter, forms the proposed NLAEC system. During training, the NN and the linear filter are jointly optimized to learn the NN parameters. In testing, the NN is used for inference and is not updated, while the linear filter is adapted to the time-varying acoustic paths.

This system requires 17 thousand parameters that consume 500 Million floating-point operations per second (Mflops) and 40 Kilo-bytes (KB) of memory, which renders it applicable for embedding on hands-free communication devices. It also meets the timing requirements of the AEC challenge \cite{Culter2021Interspeech}, and more generally the constraints of hands-free communication standards \cite{handsfree} on a standard neural processor.

Performance is evaluated against two recent NN-based NLAEC methods in \cite{malek2016hammerstein} and \cite{halimeh2019neural}, and to a linear AEC method. Experiments are conducted with 280~h of both synthetic and real data, which include half-duplex and full-duplex periods affiliated with various acoustic environments, devices, speakers, and noise and echo levels. Results show leading performance of the proposed NLAEC system in terms of echo cancellation and speech distortion levels, generalization and stability to various setups, robustness to high levels of noise and echo, and convergence and re-convergence rates.

The reminder of this paper is organized as follows. In Section \ref{ProblemFormulation}, we formulate the problem. In Section \ref{algorithm}, we introduce the proposed NLAEC system. In Section \ref{Setup}, we describe the experimental setup. In Section \ref{results}, we demonstrate the performance of the proposed system. Finally, we conclude in Section \ref{conclusion}.
\begin{figure}[t]
	\centering
	\includegraphics[width=\linewidth]{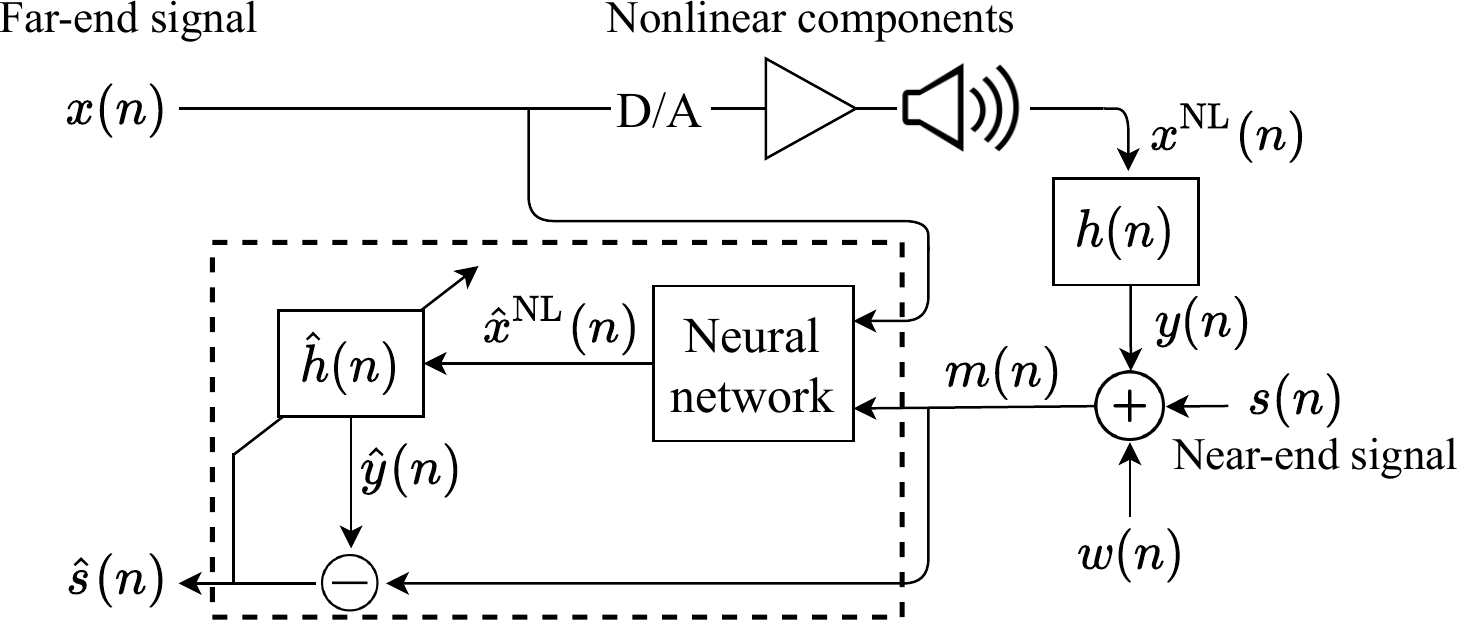}
	\caption{NLAEC scenario and proposed system (bordered). The nonlinear components are modeled with a neural network and the acoustic path with a standard adaptive linear filter.}
	\label{fig:setup}
\end{figure}
\section{Problem Formulation}\label{ProblemFormulation}
Figure~\ref{fig:setup} depicts the scenario and proposed system for NLAEC. Let $s\left(n\right)$ be the near-end speech signal and let $x\left(n\right)$ be the far-end speech signal. The microphone signal $m\left(n\right)$ is given by
\begin{align}
m\left(n\right) = s\left(n\right) + y\left(n\right) + w\left(n\right), 
\label{eq:micModel}
\end{align}
\noindent 
where $w\left(n\right)$ represents additive environmental and system noises and $y\left(n\right)$ is a nonlinear reverberant echo that is generated from  $x\left(n\right)$. The far-end signal, $x\left(n\right)$, is first distorted by electrical components that produce $x^{\textrm{NL}}\left(n\right)$, and then $x^{\textrm{NL}}\left(n\right)$ propagates via a linear acoustic path $h\left(n\right)$, namely ${y\left(n\right) = x^{\textrm{NL}}\left(n\right)\ast h\left(n\right)}$.
The proposed NLAEC system attempts to estimate $y\left(n\right)$ by  using an NN to find  $\hat{x}^{\textrm{NL}}\left(n\right)$, which is an estimate for ${x}^{\textrm{NL}}\left(n\right)$, and  filtering the result with an adaptive linear filter that tracks the acoustic path, denoted by $\hat{h}\left(n\right)$:
\begin{align}
\hat{y}\left(n\right) = \hat{x}^{\textrm{NL}}\left(n\right)\ast\hat{h}\left(n\right).
\label{eq:echoEst}
\end{align}
The signal transmitted to the far-end is given by 
\begin{align}
\hat{s}\left(n\right) = m\left(n\right) - \hat{y}\left(n\right) = s\left(n\right) + \left(y\left(n\right)-\hat{y}\left(n\right)\right) + w\left(n\right).
\label{eq:errorEst}
\end{align}
Our goal is to cancel the echo $y\left(n\right)$ by eliminating the term $y\left(n\right)-\hat{y}\left(n\right)$, without distorting the speech signal $s\left(n\right)$. 
\section{Nonlinear Acoustic Echo Cancellation}\label{algorithm}
\begin{figure}[t!]
	\centering
	\includegraphics[width=\linewidth]{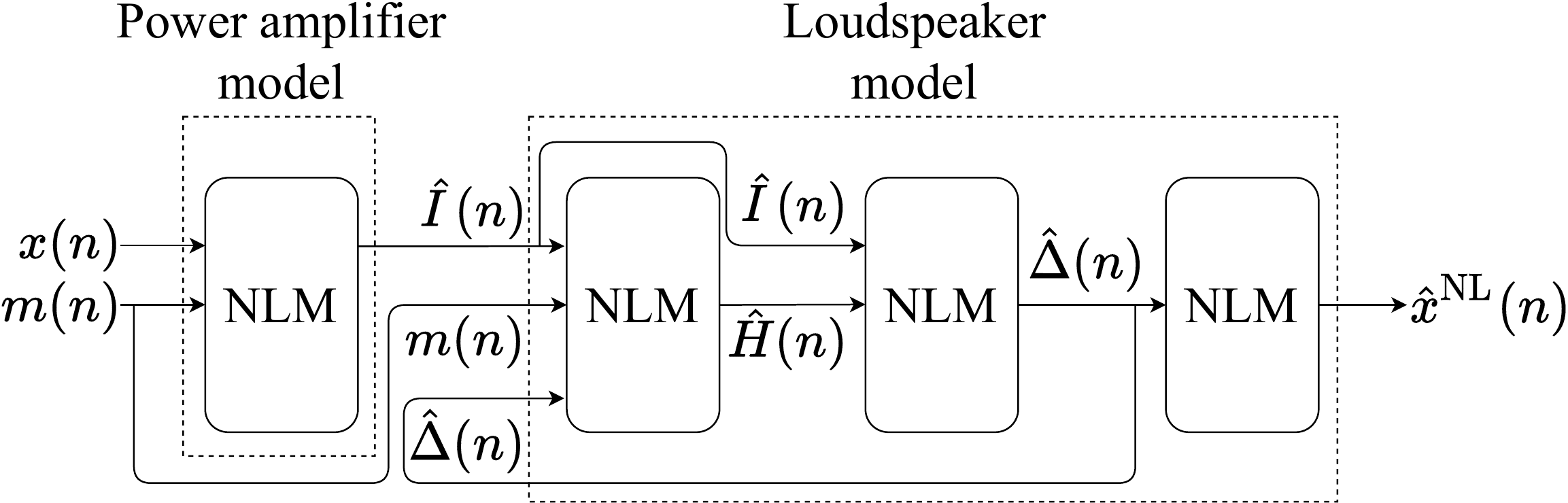}
	\caption{Proposed neural network architecture.}
	\label{fig:architecture}
\end{figure}
The proposed NLAEC system is comprised of two parts. First, an NN models the physical behavior of distortions applied between the far-end signal and the loudspeaker output, caused by non-ideal electrical components in practical hands-free devices. Second, a standard adaptive linear filter tracks the acoustic echo path from the loudspeaker output to the microphone. 

In order to understand our system, it is helpful to understand how the above-mentioned electrical components behave. Modern hands-free devices often apply distortions between receiving the far-end signal and playing it in the near-end. These distortions are created by three different electrical components; a digital-to-analog converter (D/A), a power amplifier, and a loudspeaker \cite{dobrucki2011nonlinear}, \cite{klippel2005loudspeaker}, \cite{ravaud2010ranking}, \cite{soria2004modeling}. 
This study uses a 16-bit data precision, so the signal-to-quantization-noise ratio is sufficiently high and the D/A distortions are numerically negligible \cite{dobrucki2011nonlinear}. Thus, the D/A is not modeled.
Ideally, the power amplifier should increase the energy of its input signal without distortions by using the power supply from the device battery. However, low-powered hands-free devices drive the amplifier to operate close to saturation, which yields distortions. The specific NL behavior of each amplifier depends on its saturation curve, ranging from a soft-clipped sigmoid, to a hard-clipped rectified function, and in extreme cases, it may exhibit a square waveform behavior \cite{dobrucki2011nonlinear}. 

The loudspeaker component is responsible for the majority of distortions. In this study, the widely-used electro-dynamic loudspeaker model is considered, which exhibits four major types of nonlinearities; electrical, magnetic, mechanical, and acoustical \cite{soria2004modeling}. The electrical signal, $I\left(n\right)$, is received from the amplifier output and creates a magnetic field signal of strength $H\left(n\right)$ around the voice coil, which renders it an electromagnet. The relation between $I\left(n\right)$ and $H\left(n\right)$ is NL and depends on the coil displacement signal, $\unslant{\Delta}\left(n\right)$. Both $I\left(n\right)$ and $H\left(n\right)$ lead to polarity changes in the electromagnet that moves the coil back and forth with force that also has NL relations with $\unslant{\Delta}\left(n\right)$. This movement creates air pressure that is translated into acoustic sound waves that depend on $\unslant{\Delta}\left(n\right)$ and its temporal derivatives. This relation is NL as well due to wave propagation and mechanical nonlinearities, caused by stiffness of the loudspeaker spider. 
Both the power amplifier and loudspeaker components may depend on previous observations.

The above NL behavior is modeled using an NN that is comprised of two cascaded parts: a power amplifier model, and a loudspeaker model, depicted in Figure~\ref{fig:architecture}.  
First, the amplifier is modeled with $3$ identical gated recurrent units (GRUs) that contain $16$ cells each \cite{chung2014empirical} and dropout \cite{srivastava2014dropout} in the recurrent layers, an FCNN with a one-neuron output, and a piecewise linear unit activation function layer (PLU) with trainable parameters \cite{nicolae2018plu}. This entire NL model (NLM) is fed with the far-end and microphone waveform signals, since the latter contains information about the distortions of the former.
Second, the loudspeaker is modeled by a sequence of $3$ consecutive NLMs. It receives the output of the amplifier, i.e., the estimated excitation current $\hat{I}\left(n\right)$ that drives the loudspeaker. Similarly to the amplifier model, $\hat{I}\left(n\right)$ is concatenated to the microphone signal, and the first NLM learns the electrical-to-magnetic NL model from $\hat{I}\left(n\right)$ to $\hat{H}\left(n\right)$. Then, the predicted $\hat{H}\left(n\right)$ is concatenated to $\hat{I}\left(n\right)$ and inserted to the second NLM, which learns the magnetic-to-mechanical NL model and predicts $\hat{\unslant{\Delta}}\left(n\right)$. Then, $\hat{\unslant{\Delta}}\left(n\right)$ is inserted to the third NLM, which learns the mechanical-to-acoustic NL model and estimates the distorted far-end signal at the output of the loudspeaker, i.e., $\hat{x}^{\textrm{NL}}\left(n\right)$. Since $\hat{\unslant{\Delta}}\left(n\right)$ also affects $\hat{H}\left(n\right)$, the first NLM is fed with the output of the second NLM using a skip-connection. The NLM unit is adjusted to receive between $1$ to $3$-dimensional input signals across the NN model. Following this NN, a linear adaptive filter models the acoustic path between the loudspeaker output and the microphone. This filter contains $150$ samples and was developed by Phoenix Audio Technologies\textsuperscript{TM} using a filter bank approach. The NN and the linear filter construct the proposed end-to-end NLAEC system. 

To the best of our knowledge, the proposed NN architecture is used in this study for the first time. The NN is based on the GRU, whose internal gate-based mechanism is optimized for NL sequence-to-sequence mapping in the waveform domain. Also, the GRU keeps relevant past information without discarding it through time, while neglecting irrelevant data. Thus, the optimal memory length is implicitly learned by the NN during training and should not be set in advance. The trainable PLU parameters are also adjusted during training to optimally describe various saturation curves of the power amplifier and other NL behaviors exhibited by the loudspeaker. Thus, the NL behavior of the NN is not restricted to a predetermined set of NL basis functions. In addition, the GRU consumes low computational resources and requires short inference time.

The NLAEC system contains 17 thousand parameters that consume 500 Mflops and 40 KB of memory. Thus, its integration on hands-free devices is enabled, e.g., using the NDP$120$ neural processor by Syntiant\textsuperscript{TM} \cite{Syntiant}. Timing constraints of hands-free communication on that processor are also met \cite{handsfree}. 
\section{Experimental Setup}\label{Setup}
\subsection{Database Acquisition}\label{database}
Two data corpora are employed in this study; the AEC challenge database \cite{Culter2021Interspeech}, and a database recorded in our lab, both sampled at $16$ kHz. These corpora include single-talk and double-talk periods both with and without echo-path change. In the case of no  echo-path change, there is no movement in the room during the recording. In the other case, either the near-end speaker or the device are constantly moving during the recording. In \cite{Culter2021Interspeech}, two open sources of synthetic and real recordings are introduced. The synthetic data includes $100$~h, and the real data contains $140$~h of audio clips, generated from $5,000$ hands-free devices that are used in various acoustic environments. In both real and synthetic cases, signal-to-echo ratio (SER) and signal-to-noise ratio (SNR) levels were distributed on $\left[-10,10\right]$ dB and $\left[0,40\right]$ dB, respectively. 
Additional real recordings were conducted in our lab to test the generalization of the system to unseen setups and its robustness to extremely low levels of SERs. This database is fully described in \cite{Ivry}. For completion, it contains 40~h of recordings from the TIMIT \cite{TIMIT_correct} and LibriSpeech \cite{panayotov2015librispeech} corpora with SNR levels of $32\pm5$ dB and SER levels distributed on $\left[-20,-10\right]$ dB.

Formally, the SER and SNR captured by the microphone are defined as {SER=$10\log_{10}\left[\Vert s\left(n\right) \Vert_{2}^{2} / \Vert y\left(n\right) \Vert_{2}^{2}\right]$} and {SNR=$10\log_{10}\left[\Vert s\left(n\right) \Vert_{2}^{2} / \Vert w\left(n\right) \Vert_{2}^{2}\right]$} in dB, and are calculated with 50\% overlapping time frames of 20 ms. 

\subsection{Data Processing, Training, and Testing}\label{preproc}
The real and synthetic data from \cite{Culter2021Interspeech} is randomly split to create 185~h of training set and 45~h of validation set. The test set contains only real data that is comprised of the remaining 10~h from \cite{Culter2021Interspeech} and all 40~h from \cite{Ivry}. Each set is divided into 10~s segments that contain recordings in different setups. This leads to frequent re-convergence during transitions between segments, both without and with echo-path change. These sets are balanced to prevent bias in results, as detailed in \cite{Ivry}.

During training, the NN and the succeeding linear filter are jointly optimized to learn the NN parameters. Optimization is done by minimizing the $\ell_{2}$ distance between the output of the NLAEC, $\hat{s}\left(n\right)$, and the desired-near-end speech $s\left(n\right)$.
\noindent To train the NN, back-propagation through time is used with a learning rate of 0.0005, mini-batch size of 32~ms, and 20 epochs, using Adam optimizer \cite{Adam_correct}. Also, automatic differentiation \cite{paszke2017automatic} is applied, since the loudspeaker modeling involves temporal derivatives of its input signals. Training duration was typically 15 minutes per 10~h of data on an Intel Core i7-8700K CPU @ 3.7 GHz with two GPUs of type Nvidia GeForce RTX 2080 Ti.

During testing, the NN is used for inference only and is not updated. The linear filter receives the outputs of the NN and is continuously adapted to account for time variations of the acoustic path. An artificial gain may be introduced by the NN, which is compensated as shown in \cite{vincent2006performance}. 
\begin{table}
	\centering
	\caption{Performance metrics for NLAEC.}
	\label{table:metrics}
	\renewcommand{\arraystretch}{1.5}
	\begin{tabular}{|c c|}
		\hline
		Measure & Definition \\ [0.5ex]
		\hline\hline
		ERLE &
		$10\log_{10}\frac{\Vert m\left(n\right)\Vert_{2}^{2}}{\Vert \hat{s}\left(n\right)\Vert_{2}^{2}} \Bigr|_{\textrm{Far-end single-talk}}$  \\ 
		\hline
		SDR &
		$10\log_{10}\frac{\Vert s\left(n\right)\Vert_{2}^{2}}{\Vert \hat{s}\left(n\right)-s\left(n\right)\Vert_{2}^{2}} \Bigr|_{\textrm{Double-talk}}$  \\ 
		\hline
	\end{tabular}
\end{table}

\subsection{Performance Measures}\label{measures}
To evaluate performance, the echo return loss enhancement (ERLE) \cite{ERLE} is used. It measures echo reduction between the degraded and enhanced signals when only a far-end signal and noise are present. For double-talk periods, we use the signal-to-distortion ratio (SDR) \cite{vincent2006performance} that takes echo suppression and speech distortion into account, and the perceptual evaluation of speech quality (PESQ) \cite{PESQ}, \cite{PESQ_wb_ext}. The PESQ is calculated over an entire 10~s segment. The ERLE and SDR are calculated with 50\% overlapping frames of 20~ms, and are defined in Table \ref{table:metrics}.

\section{Experimental Results}\label{results}
The performance of the proposed NLAEC system is compared against two competing NN-based methods in \cite{malek2016hammerstein} and \cite{halimeh2019neural}, notated ``Malek'' and ``Halimeh'', respectively. To approximate the linear echo path, the proposed system and ``Malek'' are implemented here with an identical adaptive linear filter mentioned in Section \ref{algorithm}, while ``Halimeh'' employs a linear echo approximation via an NN. As benchmark, the linear filter is also applied alone, and this method is denoted by ``Linear''. Measures are reported by their mean and standard deviation (std) values, with respect to the test set specified in each experiment. Unless stated otherwise, the format of the results is presented as mean$\pm$std. In this study, convergence was reached if the normalized misalignment between consecutive linear echo approximations was lower than $-30$ dB \cite{paleologu2015overview}.  

Results for segments with no echo-path change are given in Table~\ref{table:aecNoEchoChange} and for segments with echo-path change are given in Table~\ref{table:aecEchoChange}, both after convergence. Compared to competition, the proposed method achieves enhanced echo cancellation in single-talk periods according to the ERLE measure. In double-talk periods, less speech distortion and better speech quality are obtained, as suggested by the SDR and PESQ scores, respectively. Also, a lower std measure is achieved, which projects better stability of our method across various setups.
Scenarios of echo-path change lead to overall decline in performance relative to no echo-path change, as expected. However, our method still prevails competition across all measures in terms of both higher mean and lower std. 
Based on the above, our method allows enhanced modeling of the NL echo path, which improves both the estimation of acoustic paths with no echo-path change, and the tracking of acoustic paths with echo-path change.

In addition, we investigate the performance before convergence and during re-convergence for segments with no echo-path change. Due to the test set segmentation described in Section \ref{preproc}, re-convergence frequently occurs during transitions between segments. As shown in Table \ref{table:aecBeforeConvergence}, performance is collectively impeded relative to the converged case in Table~\ref{table:aecNoEchoChange}. However, our method still prevails across all measures in terms of both mean and std values. This indicates the high sensitivity of competing methods to converged echo approximation, while our model captures the behavior of the echo even from degraded measurements. We also examine the convergence time of each method. According to Table \ref{table:aecConvergenceTimes}, our method achieves the shortest convergence time compared to competition. Again, it can be suggested that enhanced modeling of the NL echo path is obtained by the proposed NN, which allows the succeeding linear filter to be adjusted more accurately and rapidly.

\begin{table}[t!]
	\small
	\renewcommand{\arraystretch}{1.1}
	\caption{Performance with no echo-path change.}
	\label{table:aecNoEchoChange}
	\centering
	\begin{tabular}{P{0.04\textwidth}P{0.07\textwidth}P{0.07\textwidth}P{0.07\textwidth}P{0.07\textwidth}} 
		\multicolumn{1}{c}{}
		& \multicolumn{1}{c}{Proposed}
		& \multicolumn{1}{c}{Halimeh}
		& \multicolumn{1}{c}{Malek}
		& \multicolumn{1}{c}{Linear}
		\\
		\hline
		ERLE
		&  \textbf{26.4$\pm$5.1} &  23.1$\pm$5.9  &  22.6$\pm$6.7  &  21.3$\pm$7.2  \\
		\hline
		PESQ
		&  \textbf{3.17$\pm$0.4} &  2.88$\pm$0.5  &  2.64$\pm$0.5  &  2.02$\pm$0.7 \\
		\hline
		SDR
		&  \textbf{5.37$\pm$0.4} &  4.83$\pm$0.6  &  4.37$\pm$0.8  &  3.01$\pm$0.9  \\
		\hline
	\end{tabular}
\end{table}
\begin{table}[t!]
	\small
	\renewcommand{\arraystretch}{1.1}
	\caption{Performance with echo-path change.}
	\label{table:aecEchoChange}
	\centering
	\begin{tabular}{P{0.04\textwidth}P{0.07\textwidth}P{0.07\textwidth}P{0.07\textwidth}P{0.07\textwidth}} 
		\multicolumn{1}{c}{}
		& \multicolumn{1}{c}{Proposed}
		& \multicolumn{1}{c}{Halimeh}
		& \multicolumn{1}{c}{Malek}
		& \multicolumn{1}{c}{Linear}
		\\
		\hline
		ERLE
		&  \textbf{23.2$\pm$6.0} &  19.2$\pm$7.7  &  18.0$\pm$8.3  &  16.9$\pm$8.9  \\
		\hline
		PESQ
		&  \textbf{2.92$\pm$0.5} &  2.54$\pm$0.7  &  2.31$\pm$0.6  &  1.91$\pm$0.6 \\
		\hline
		SDR
		&  \textbf{5.08$\pm$0.6} &  4.25$\pm$0.9  &  3.82$\pm$0.9  &  2.52$\pm$1.0  \\
		\hline
	\end{tabular}
\end{table}
\begin{table}[t!]
	\small
	\renewcommand{\arraystretch}{1.1}
	\caption{Performance before convergence.}
	\label{table:aecBeforeConvergence}
	\centering
	\begin{tabular}{P{0.04\textwidth}P{0.07\textwidth}P{0.07\textwidth}P{0.07\textwidth}P{0.07\textwidth}} 
		\multicolumn{1}{c}{}
		& \multicolumn{1}{c}{Proposed}
		& \multicolumn{1}{c}{Halimeh}
		& \multicolumn{1}{c}{Malek}
		& \multicolumn{1}{c}{Linear}
		\\
		\hline
		ERLE
		&  \textbf{19.7$\pm$7.5} &  14.9$\pm$8.1  &  13.8$\pm$8.8  &  11.0$\pm$9.6  \\
		\hline
		PESQ
		&  \textbf{2.56$\pm$0.6} &  1.98$\pm$0.7  &  1.91$\pm$0.7  &  1.75$\pm$0.6 \\
		\hline
		SDR
		&  \textbf{4.71$\pm$0.9} &  3.58$\pm$1.2  &  3.04$\pm$1.3  &  1.54$\pm$1.3 \\
		\hline
	\end{tabular}
\end{table}
\begin{table}[t!]
	\small
	\renewcommand{\arraystretch}{1.1}
	\caption{Convergence time in seconds.}
	\label{table:aecConvergenceTimes}
	\centering
	\begin{tabular}{m{0.07\textwidth}m{0.07\textwidth}m{0.07\textwidth}m{0.07\textwidth}} 
		\multicolumn{1}{c}{Proposed}
		& \multicolumn{1}{c}{Halimeh}
		& \multicolumn{1}{c}{Malek}
		& \multicolumn{1}{c}{Linear}
		\\
		\hline
		\textbf{4.6$\pm$0.7} &  6.6$\pm$1.1  &  7.3$\pm$1.4  &  7.9$\pm$1.8 \\
		\hline
	\end{tabular}
\end{table}
\begin{figure}
	\centering
	\begin{subfigure}[t]{0.23\textwidth}
		\centering
		\includegraphics[width=0.93\textwidth]{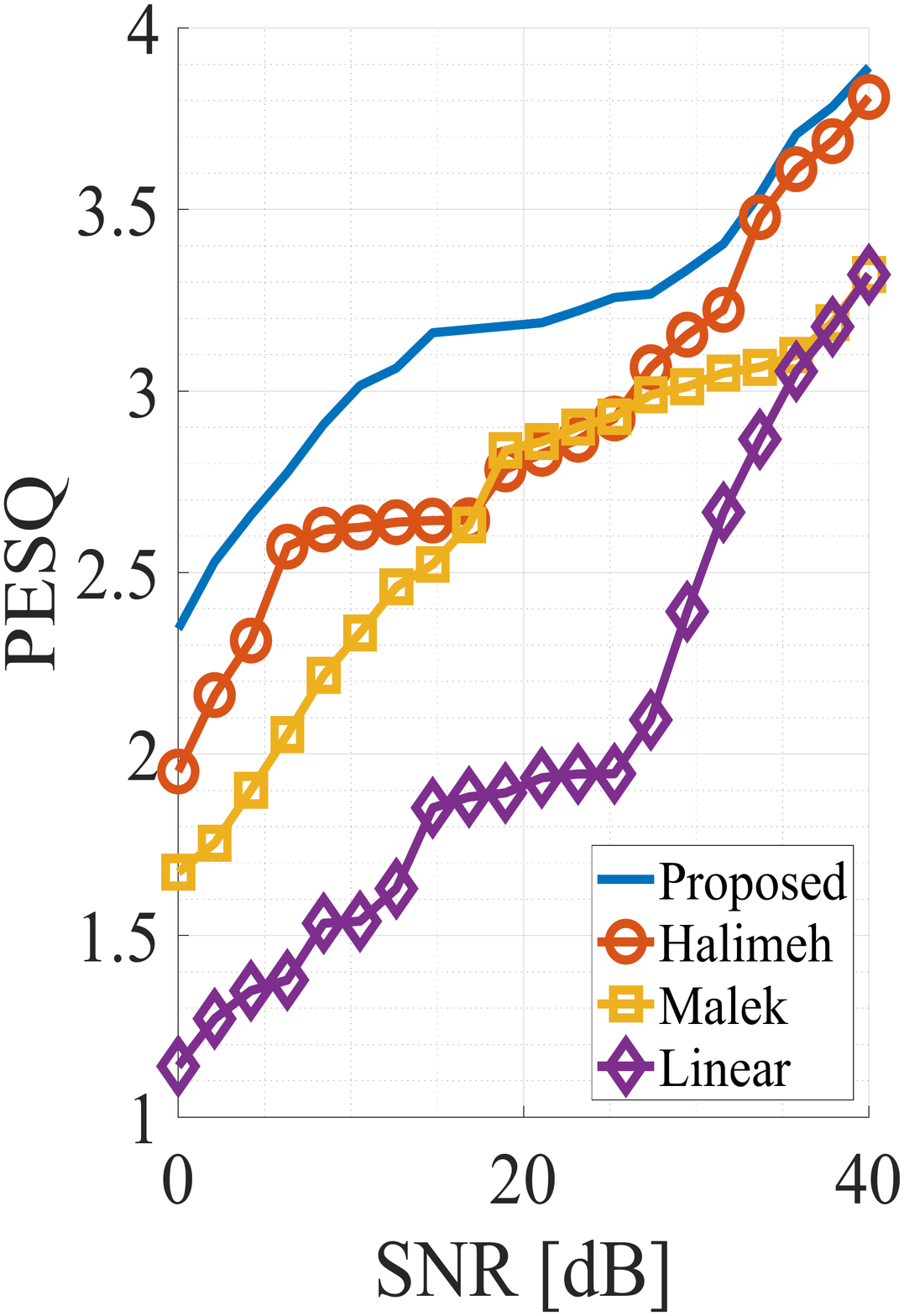}
	\end{subfigure}
	\hfill
	\begin{subfigure}[t]{0.23\textwidth}
		\centering
		\includegraphics[width=0.93\textwidth]{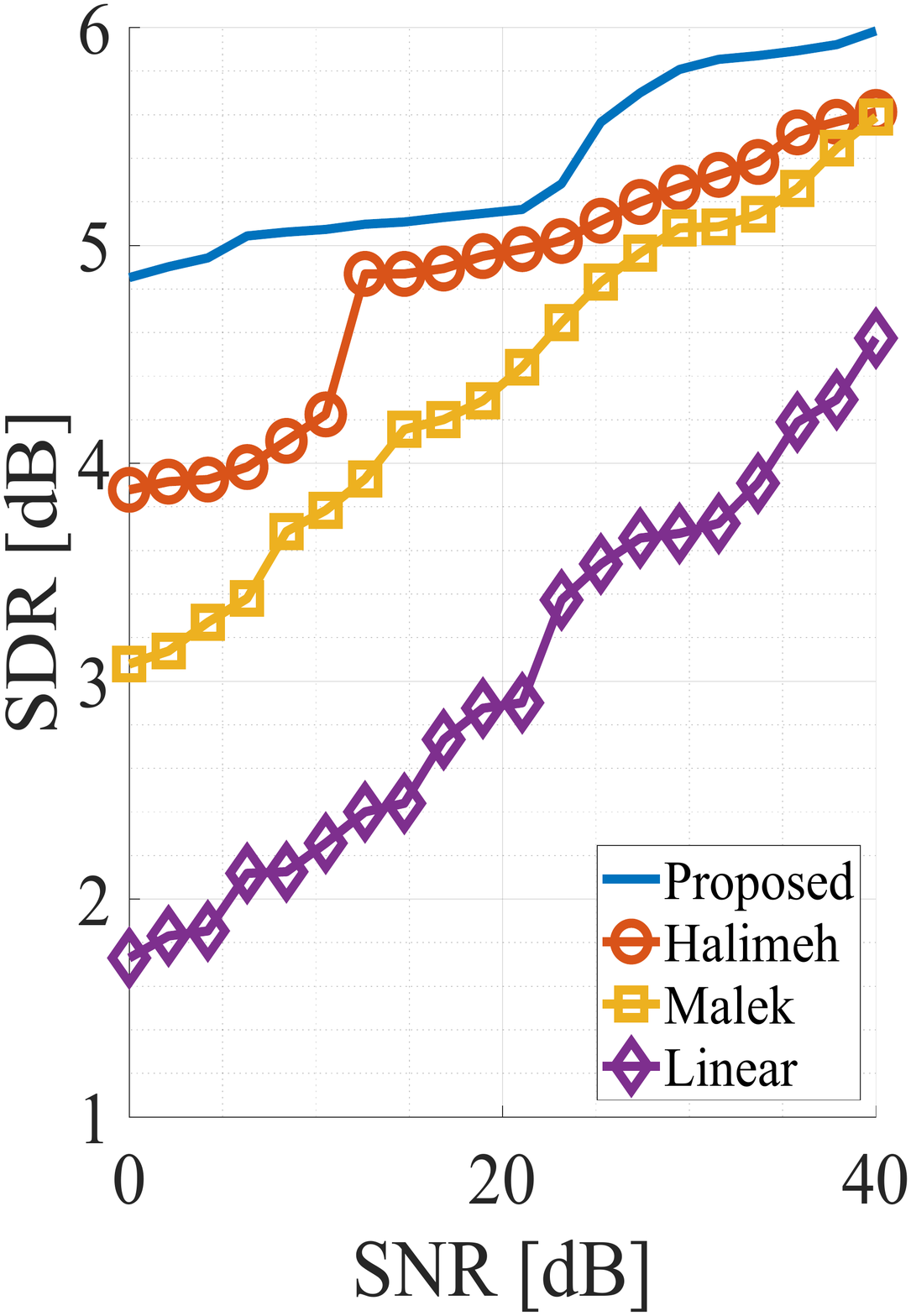}
	\end{subfigure}
	
	\begin{subfigure}[t]{0.23\textwidth}
		\centering
		\includegraphics[width=0.93\textwidth]{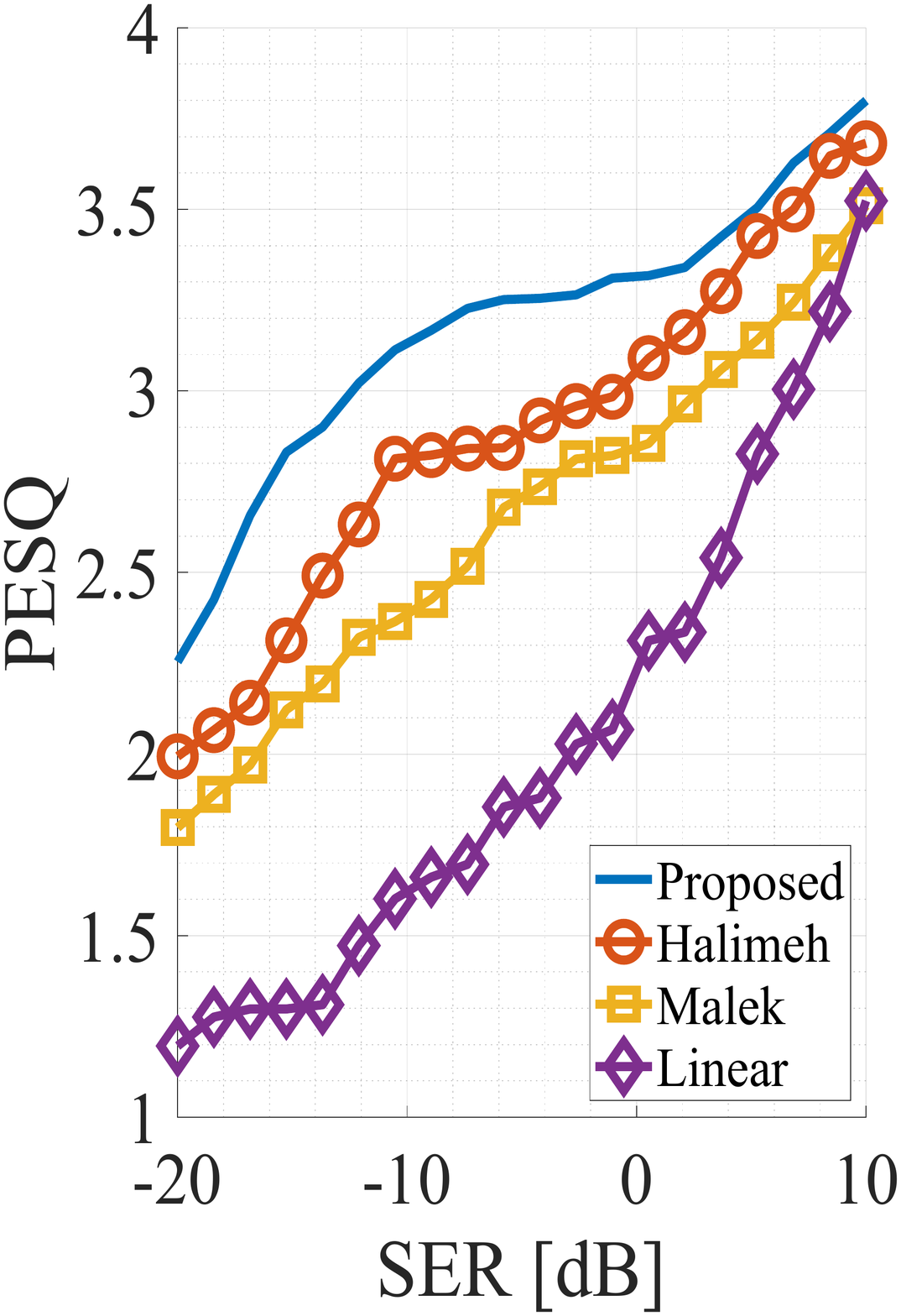}
	\end{subfigure}
	\hfill
	\begin{subfigure}[t]{0.23\textwidth}
		\centering
		\includegraphics[width=0.93\textwidth]{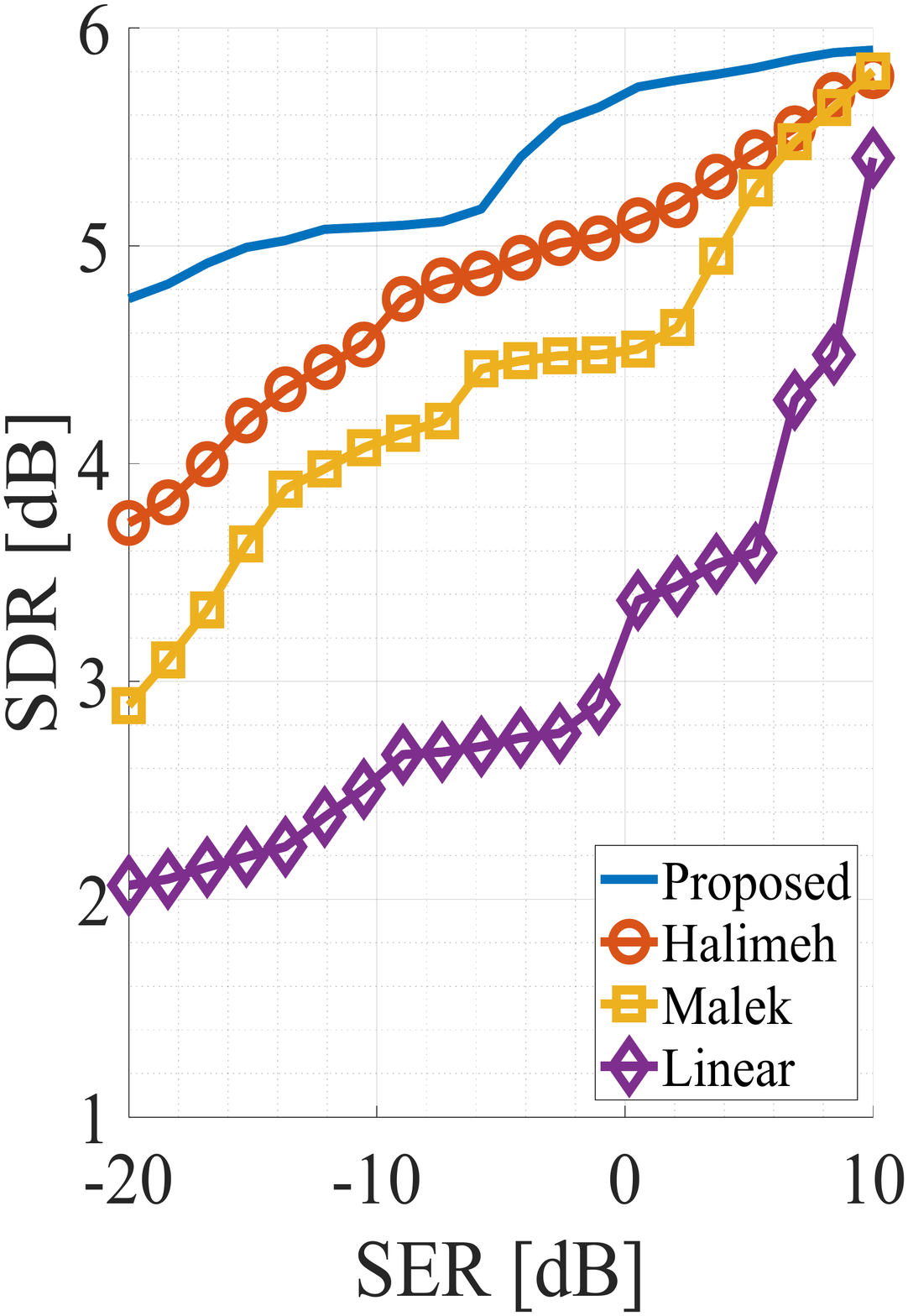}
	\end{subfigure}
\caption{Comparison of average PESQ and SDR measures in various SNR and SER levels.}
\label{fig:snr_ser}
\end{figure}

Next, performance with no echo-path change is examined in various SNR and SER levels, after convergence. As shown in Figure~\ref{fig:snr_ser}, all methods suffer from decline in performance when acoustic conditions deteriorate. However, our method outperforms competition in both PESQ and SDR measures across all SNR and SER levels, which projects high generalization ability to various levels of noise and echo. The relatively stable behavior of the proposed method, especially in low levels of SNRs and SERs, indicates high robustness to high levels of noise and echo that often occur in practice. 
Interestingly, in severely degraded conditions of $0$ dB SNR and of $-20$ dB SER, the proposed method achieves roughly 1 dB higher SDR and 0.5 higher PESQ score on average than the competition in second place.

\section{Conclusion}
\label{conclusion}
We have presented an NLAEC system that comprises a novel NN architecture and a succeeding standard adaptive linear filter. To describe the distortions modern hands-free devices induce between receiving and playing the far-end signal, we constructed the NN of a power amplifier model followed by a loudspeaker model. The adaptive filter is fed by the NN and tracks the acoustic path from the loudspeaker output to the microphone. The NN parameters are  updated during training using joint optimization of the NN and the  filter.
The NLAEC implementation is adequate for integration on hands-free devices, and can meet timing requirements of hands-free communication standards on a standard neural processor.
Experiments with 280 h of real and synthetic recordings demonstrate the improved performance of our method compared to competition in terms of echo suppression and desired-signal distortion, generalization and stability in various setups, robustness to high levels of noise and echo, and convergence and re-convergence times. 


\bibliographystyle{IEEEtran}
\bibliography{NLAEC_refs}

\end{document}